\documentclass[showpacs,amsmath,amssymb]{revtex4}

\usepackage{graphicx}
\usepackage{dcolumn}
\usepackage{bm}
\usepackage{epsf}
\usepackage{feynmp}

\newcommand{\setval}{\fmfset{wiggly_len}{1.5mm}
\fmfset{arrow_len}{1.5mm}\fmfset{arrow_ang}{13}\fmfset{dash_len}{1.5mm}
\fmfpen{0.125mm}\fmfset{dot_size}{1thick}}
\fmfset{dot_size}{1thick}
\begin{document}
\begin{fmffile}{putz}
\title{
Variational Perturbation Theory for Markov Processes}
\author{Hagen Kleinert, Axel Pelster, and Mihai V. Putz\\
       Institut f\"ur Theoretische Physik, Freie Universit\"at Berlin,
        Arnimallee 14, D--14195 Berlin, Germany\\ 
E-mails: kleinert@physik.fu-berlin.de, pelster@physik.fu-berlin.de, putz@physik.fu-berlin.de\\}
\date{\today}

\begin{abstract}
We develop a convergent variational perturbation theory for conditional probability densities of Markov processes.
The power of the theory is illustrated by applying it to the diffusion of a particle in an anharmonic potential.
\end{abstract}
\maketitle
\section{Introduction}
Variational perturbation theory \cite{Kleinert} transforms divergent perturbation expansions into convergent ones with
the convergence extending to infinitely strong couplings. The theory has been developed 
for the path integral representation of the free 
energy and the density matrix in quantum statistics, and has been tested for many systems, in particular
the anharmonic oscillator and the hydrogen 
atom without and with a homogeneous magnetic field
\cite{Kleinert,Janke,Meyer,Kuerzinger,Bachmann1,Bachmann2,Bachmann3}. The
procedure is based on approximating a potential by a local trial oscillator whose frequency is optimized
order by order for each set of external end points. Recently, variational perturbation theory has also been successfully extended
to statistical field theory to calculate highly accurate critical properties of second-order phase transitions \cite{Verena}.\\

In this paper we develop a similarly convergent variational perturbation theory for the path integral
representation of the conditional probability density of Markov processes. In close analogy with the previous method 
we approximate a given
stochastic process by a trial Brownian motion with a linear drift coefficient, and optimize the 
damping constant. We illustrate the procedure by calculating the
time dependence of the conditional probability density for a nonlinear stochastic model. 
After some introductory remarks on Markov processes in Section II,
the path integral for the conditional probability density 
is treated perturbatively in Section III, and 
evaluated variationally in Section IV.
\section{Markov Processes}
We start by summarizing the basic properties of Markov processes \cite{Stratonovich,Kampen,Haken1,Risken}
needed in the sequel.
\subsection{Fokker-Planck Equation}
A Markov process for a single stochastic variable $x$ is characterized 
by the conditional probability density $P(x_b\, t_b | x_a\, t_a)$ that 
the event $x_b$ is realized at the time $t_b$ once the event $x_a$ has 
taken place at time $t_a$. It has the initial condition
\begin{equation}
\label{IVP}
P (x_b\, t_a | x_a\, t_a) = \delta(x_b-x_a) 
\end{equation}
and obeys, for additive noise, the Fokker-Planck equation 
\begin{equation}
\label{FPE}
\frac{\partial }{\partial t_b} 
P (x_b\, t_b | x_a\, t_a) 
= \hat{L}_{\rm FP} (x_b ) \,
P (x_b\, t_b | x_a\, t_a) \, ,
\end{equation}
with the Fokker-Planck operator
\begin{equation}
\label{FPO}
\hat{L}_{\rm FP} (x_b) \bullet = -
\frac{\partial}{\partial x_b}\, \left[ K(x_b) \bullet \right] 
+ D\, \frac{\partial^2}{\partial x_b^2} \bullet \, , 
\end{equation}
where $K(x)$ and $D$ denote the drift and diffusion coefficient, 
respectively. 
An important  example is provided by 
the overdamped motion of a Brownian particle with mass $M$ and 
friction constant $\kappa$ in an external potential  $V(x)$. In this 
case, the drift coefficient reads
\begin{equation}
\label{BM1}
K(x) = - \frac{V' ( x )}{M \kappa} \, ,
\end{equation}
and the diffusion coefficient is proportional to the temperature $T$ via Einstein's relation
\begin{equation}
\label{BM2}
D = M \kappa k_B T \, .
\end{equation}
Since the spatial derivatives in the Fokker-Planck equation (\ref{FPE}), (\ref{FPO}) are all on the left-hand side,
they guarantee the probability conservation
\begin{eqnarray}
\frac{\partial}{\partial t_b} \int_{- \infty}^{+ \infty} d x_b
P(x_b\, t_b | x_a\, t_a) = 0 \, ,
\end{eqnarray}
such that the normalization integral, which is unity at the initial time $t_b=t_a$
due to (\ref{IVP}), remains so for all times:
\begin{eqnarray}
\label{NORM}
\int_{- \infty}^{+ \infty} d x_b P(x_b\, t_b | x_a\, t_a) = 1 \, .
\end{eqnarray}
In the long-time limit $t_b \rightarrow \infty$, the conditional probability density
$P(x_b\, t_b | x_a\, t_a)$ becomes {\em stationary}
\begin{equation}
\label{SS1}
\lim_{t_b \rightarrow \infty}
P(x_b\, t_b | x_a\, t_a) = P_{\rm st}(x_b),
\end{equation}
where $P_{\rm st} (x_b)$ denotes the time-independent solution 
of the Fokker-Planck equation (\ref{FPE}), defined by
\begin{equation}
\hat{L}_{\rm FP} (x_b) P_{\rm st} (x_b) = 0 \, .
\end{equation}
By applying the Fokker-Planck operator (\ref{FPO}), we verify this solution is
\begin{equation}
\label{SS2}
P_{\rm st}(x_b) = N \exp \left[ \frac{1}{D} \int^{x_b} dx \, K(x) 
\, \right]\, ,
\end{equation}
where the normalization constant $N$ follows from (\ref{NORM}) and (\ref{SS1}) 
\begin{equation}
\label{SS3}
N = \left\{ \int^{+\infty}_{-\infty} dx_b \exp \left[ 
\frac{1}{D} \int^{x_b} dx \, K(x) \, \right] \right\}^{-1} \, .
\end{equation}
\subsection{Path Integral}
The solution of the Fokker-Planck initial value problem (\ref{IVP})--(\ref{FPO}) 
has the path integral representation
\begin{equation}
\label{PI1}
P(x_b\, t_b | x_a\, t_a) =
\int^{x(t_b)=x_b}_{x(t_a)=x_a} 
{\cal D} x e^{- {\cal A}[x]} 
\end{equation}
with the generalized Onsager-Machlup functional
\begin{eqnarray}
\label{PI2}
{\cal A} [x] = \frac{1}{4 D}\int^{t_b}_{t_a} dt [\dot{x} ( t ) - K(x(t)) ]^2 + \frac{1}{2} \int^{t_b}_{t_a} dt K' ( x(t) ) \, ,
\end{eqnarray}
where all paths $x(t)$ contribute which connect the spacetime points $(x_a, t_a)$ 
and $(x_b, t_b)$. The extra term in (\ref{PI2}) is needed since the path 
integral (\ref{PI1}), (\ref{PI2}) is by definition symmetrically ordered in the product of $\dot{x} ( t )$ and $K ( x ( t ) )$, 
corresponding to a midpoint discretization \cite{Kleinert}:
\begin{eqnarray}
P(x_b\, t_b | x_a\, t_a) &=& \lim_{\epsilon \rightarrow 0} \left\{ \prod_{n=1}^N \int\limits_{- \infty}^{+ \infty}
d x_n \right\} \left( \frac{1}{4 \pi D \epsilon} \right)^{\frac{N+1}{2}} \nonumber\\
&& \times \exp \left\{ - \frac{\epsilon}{4 D} \sum_{n=1}^{N+1} \left[ \frac{x_n -x_{n-1}}{\epsilon} - K \left( 
\frac{x_n + x_{n-1}}{2} \right) \right]^2 - \frac{\epsilon}{2} \sum_{n=1}^{N+1} K' \left( 
\frac{x_n + x_{n-1}}{2} \right) \right\} \, .
\end{eqnarray}
Making use of the stationary solution (\ref{SS2}), the path integral (\ref{PI1}), (\ref{PI2})
can be factorized as
\begin{eqnarray}
\label{FACTOR}
P ( x_b \, t_b | x_a \, t_a ) = \sqrt{\frac{P_{\rm st}(x_b)}{P_{\rm st}(x_a)}}
\, \tilde{P} ( x_b \, t_b | x_a \, t_a ) 
\end{eqnarray}
with the remaining path integral
\begin{eqnarray}
\label{SPI}
\tilde{P} ( x_b \, t_b | x_a \, t_a ) = \int_{x(t_a)=x_a}^{x(t_b)=x_b}
{\cal D} x \exp \left\{ - \frac{1}{2 D} \int_{t_a}^{t_b} d t
\left[ \frac{1}{2} \dot{x}^2 ( t ) + \frac{1}{2} K^2 ( x ( t ) )  
+ D K' ( x ( t ) ) \right] \right\} \, .
\end{eqnarray}
This coincides with the quantum-statistical imaginary-time
evolution amplitude \cite{Kleinert}
\begin{eqnarray}
\label{QSPI}
( x_b \, \hbar \beta | x_a \, 0 ) = \int_{x(0)=x_a}^{x(\hbar \beta)=x_b}
{\cal D} x \exp \left\{ - \frac{1}{\hbar} \int_{0}^{\hbar \beta} d \tau
\left[ \frac{M}{2} \dot{x}^2 ( \tau ) + U ( x ( \tau ) )  
\right] \right\} \, ,
\end{eqnarray}
in which we identify
\begin{eqnarray}
\label{FORMAL}
t_a \equiv 0 \, , \hspace*{1cm} t_b \equiv \hbar \beta \, , \hspace*{1cm} D \equiv \frac{\hbar}{2 M} 
\end{eqnarray}
and set
\begin{eqnarray}
U ( x ) \equiv \frac{M}{2} K^2 ( x ) + \frac{\hbar}{2} K' ( x ) \, .
\end{eqnarray}
To the path integral (\ref{QSPI}), we can apply directly the known
variational perturbation theory \cite{Kleinert}, which will lead us in the present context to a solution of
the Fokker-Planck initial value problem (\ref{IVP})--(\ref{FPO}).
\subsection{Brownian Motion}
A solvable trial path integral is provided by
the Brownian motion with a linear drift coefficient
\begin{eqnarray}
\label{BROWN}
K ( x ) = - \kappa x  \,,
\end{eqnarray}
where the stationary solution (\ref{SS2}), (\ref{SS3}) reads
\begin{eqnarray}
P_{\kappa,{\rm st}} ( x ) = \sqrt{\frac{\kappa}{2 \pi D}} \exp \left( -
\frac{\kappa}{2D} x^2 \right) \, .
\end{eqnarray}
The conditional probability density factorizes therefore according to (\ref{FACTOR}) as
\begin{eqnarray}
\label{FAC}
P_{\kappa} ( x_b \, t_b | x_a \, t_a ) = \exp \left[ \frac{\kappa}{2} ( t_b - t_a )
- \frac{\kappa}{4 D} \left( x_b^2 - x_a^2 \right) \right]
\tilde{P}_{\kappa} ( x_b \, t_b | x_a \, t_a ) \, ,
\end{eqnarray}
and the remaining path integral is simply
\begin{eqnarray}
\label{BMP}
\tilde{P}_{\kappa} 
( x_b \, t_b | x_a \, t_a ) = \int_{x(t_a)=x_a}^{x(t_b)=x_b}
{\cal D} x \exp \left\{ - \frac{1}{2 D} \int_{t_a}^{t_b} d t
\left[ \frac{1}{2} \dot{x}^2 ( t ) + \frac{1}{2} \kappa^2 x^2 ( t ) \right]
\right\} \, .
\end{eqnarray}
It describes a quantum-statistical harmonic oscillator
with the potential $U(x) = M \kappa^2 x^2/2$.
Inserting the imaginary-time evolution amplitude of the harmonic
oscillator \cite{Neu}
and taking into account the identification
(\ref{FORMAL}), we obtain
\begin{eqnarray}
\label{ABB}
\tilde{P}_{\kappa} ( x_b \, t_b | x_a \, t_a ) = \sqrt{\frac{\kappa}{4 \pi D \sinh \kappa (t_b - t_a)}}
\, \exp \left\{ \frac{-\kappa}{4 D \sinh \kappa (t_b - t_a )} \left[ \left( x_a^2 + x_b^2 \right) \cosh \kappa ( t_b - t_a )
- 2 x_a x_b \right] \right\} \, .
\end{eqnarray}
The resulting conditional probability density of the Brownian motion (\ref{BROWN}) follows from (\ref{FAC}), (\ref{ABB})
and leads to the well-known Gaussian distribution 
\begin{eqnarray}
\label{G1}
P_{\kappa} ( x_b \, t_b | x_a \, t_a ) =  
\frac{1}{\sqrt{2 \pi \sigma^2 ( x_a , t_a ; t_b )}}
\exp \left\{ - \frac{\left[ x_b - \overline{x} ( x_a , t_a ; t_b ) 
\right]^2}{2 \sigma^2 ( x_a , t_a ; t_b )} \right\}
\end{eqnarray}
with the average point
\begin{eqnarray}
\overline{x} ( x_a , t_a ; t_b ) & = & x_a e^{- \kappa ( t_b - t_a )} \, , 
\label{G2} 
\end{eqnarray}
and the width
\begin{eqnarray}
\sigma ( x_a , t_a ; t_b ) & = & \sqrt{\frac{D}{\kappa} \left[ 1 - 
e^{-2 \kappa(t_b - t_a)} \right]} \, .
\label{G3}
\end{eqnarray}
It can easily be verified that the conditional probability density (\ref{G1})--(\ref{G3}) satisfies the initial condition (\ref{IVP}),
and obeys the Fokker-Planck equation associated with the drift coefficient (\ref{BROWN}):
\begin{eqnarray}
\frac{\partial}{\partial t_b} P_{\kappa} ( x_b \, t_b | x_a \, t_a )
= \frac{\partial}{\partial x_b} \left[ \kappa x_b  
P_{\kappa} ( x_b \, t_b | x_a \, t_a )
\right] + D \frac{\partial^2}{\partial x_b^2}
P_{\kappa} ( x_b \, t_b | x_a \, t_a ) \, . 
\end{eqnarray}
In general the drift coefficient $K ( x )$ of a Markov process is a nonlinear function in $x$, and
the corresponding conditional probability density $P ( x_b \, t_b | x_a \, t_a )$ 
cannot be calculated exactly. We must then resort to approximation methods, and we want to show in this paper that
variational perturbation theory is a very efficient one.
For this we first need an ordinary perturbation expansion which will be derived in
the next section for a typical example.
\section{Perturbation Theory}
\begin{figure}[h]
\centerline{\includegraphics[width=10cm]{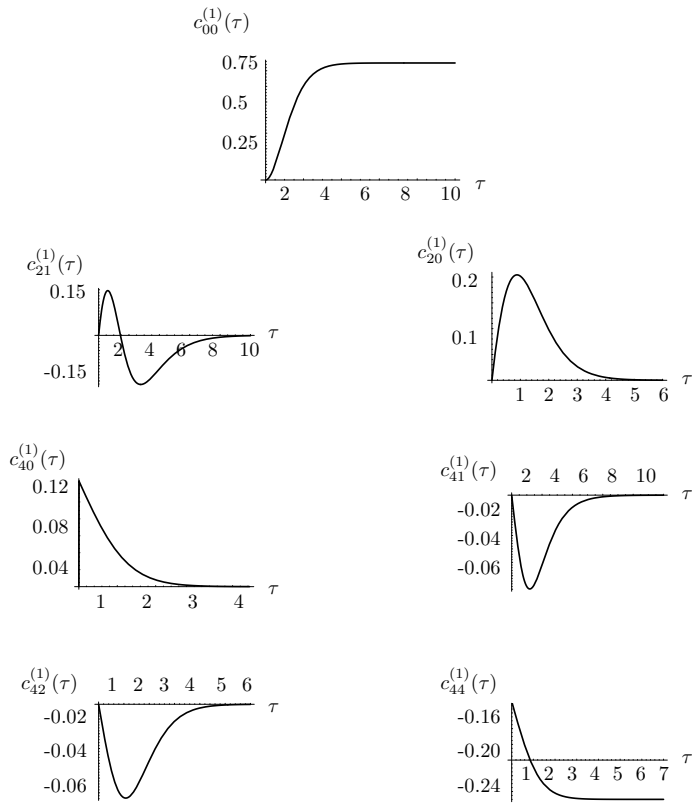}}
\caption{\label{F1}Temporal behavior of the expansion coefficients (\ref{COEFF}) of the conditional probability density (\ref{FORM})
as a function of $\tau = \kappa ( t_b - t_a )$ for $D=\kappa=1$.}
\end{figure}

Consider the nonlinear drift coefficient 
\begin{eqnarray}
\label{LASER}
K ( x ) = - \kappa x - g x^3
\end{eqnarray}
with a coupling constant $g$. Such a stochastic model is useful, for 
instance, to describe the statistical properties of laser light near
the threshold in semiclassical laser theory \cite{Risken,Haken3}.
In this case, the stochastic variable $x$ is identified with the electric
field. The parameter $\kappa$ is proportional to the difference between
the pump parameter $\sigma$ and its threshold value $\sigma_{\rm thr}$.
The coupling constant $g$ describes the interaction between light and
matter within the dipole approximation, and the diffusion constant $D$
characterizes the spontaneous emission of radiation.
For such a stochastic process, the stationary solution 
(\ref{SS2}), (\ref{SS3}) reads
\begin{eqnarray}
\label{NLST}
P_{\rm st} ( x ) = \sqrt{\frac{2g}{\kappa}} \,\, 
\frac{\exp \left[ - {\displaystyle \frac{1}{2 D} 
\left( \frac{\kappa^2}{4g}+ \kappa x^2 + \frac{g}{2} 
x^4 \right)}\right]}{{\displaystyle 
K_{1/4} \left( \frac{\kappa^2}{8 D g}\right)}} \, ,
\end{eqnarray}
where $K_{\nu} ( z )$ denotes a modified Bessel function \cite{Gradshteyn}.
The path integral for the conditional probability density corresponding to (\ref{PI1}), (\ref{PI2}) reads:
\begin{eqnarray}
P ( x_b \, t_b | x_a \, t_a ) = \int_{x(t_a)=x_a}^{x(t_b)=x_b}
{\cal D} x \exp \left\{ - \frac{1}{4D} \int_{t_a}^{t_b} d t
\left[ \dot{x} ( t ) + \kappa x ( t ) + g x^3 ( t ) \right]^2
+ \frac{1}{2} \int_{t_a}^{t_b} d t 
\left[ \kappa + 3 g x^2 ( t ) \right] \right\} \, .
\end{eqnarray}
The decomposition of type (\ref{FACTOR}) leads to
\begin{eqnarray}
\label{SMF}
P ( x_b \, t_b | x_a \, t_a ) = \exp \left[ \frac{\kappa}{2} ( t_b - t_a ) - \frac{\kappa}{4 D} \left( x_b^2 - x_a^2 \right)
- \frac{g}{8 D} \left( x_b^4 - x_a^4 \right) \right] \tilde{P} ( x_b \, t_b | x_a \, t_a ) 
\end{eqnarray}
with the remaining path integral
\begin{eqnarray}
\label{DDD}
\tilde{P} ( x_b \, t_b | x_a \, t_a ) =
\int_{x(t_a)=x_a}^{x(t_b)=x_b}
{\cal D} x \exp \left\{ - \frac{1}{2D} \int_{t_a}^{t_b} d t
\left[ \frac{1}{2} \dot{x}^2 ( t ) + \frac{1}{2} \kappa^2 
x^2 ( t ) + g \kappa x^4 ( t ) + \frac{1}{2} g^2 x^6 ( t ) - 3 g D x^2 ( t ) \right]
\right\} \, .
\end{eqnarray}
For zero coupling constant $g$, we obtain the conditional probability density
(\ref{BMP}) of the Brownian motion (\ref{BROWN}). Expanding the exponential in powers of $g$,
we find the first-order approximation
\begin{eqnarray}
\label{PE}
\tilde{P} ( x_b \, t_b | x_a \, t_a ) = 
\tilde{P}_{\kappa} ( x_b \, t_b | x_a \, t_a ) \left\{
1 + g \left[ \frac{3}{2} \int_{t_a}^{t_b} d t \, \langle x^2 ( t ) 
\rangle_{\kappa} - \frac{\kappa}{2 D} \int_{t_a}^{t_b} d t \,
\langle x^4 ( t ) \rangle_{\kappa} \right] + \, \ldots  \right\}  \, .
\end{eqnarray}
On the right-hand side, we have denoted the harmonic expectation value of any 
functional $F [ x]$ of the path $x ( t )$ by
\begin{eqnarray}
\langle F [x] \rangle_{\kappa} = \frac{1}{\tilde{P}_{\kappa} 
( x_b \, t_b | x_a \, t_a )} \int_{x(t_a)=x_a}^{x(t_b)=x_b}
{\cal D} x F [ x] \exp \left\{ - \frac{1}{2 D} 
\int_{t_a}^{t_b} d t \left[ \frac{1}{2} \dot{x}^2 ( t ) + 
\frac{1}{2} \kappa^2  x^2 ( t ) \right] \right\} \, .
\end{eqnarray}
The latter is evaluated with the help of the generating functional
for the quantum-statistical harmonic oscillator
\begin{eqnarray}
\tilde{P}_{\kappa} ( x_b \, t_b | x_a \, t_a )[j] = 
\int_{x(t_a)=x_a}^{x(t_b)=x_b}
{\cal D} x \exp \left\{ - \frac{1}{2 D} 
\int_{t_a}^{t_b} d t \left[ \frac{1}{2} \dot{x}^2 ( t ) + 
\frac{1}{2} \kappa^2  x^2 ( t ) - j ( t ) x ( t ) \right] \right\} \, ,
\end{eqnarray}
which has the explicit form \cite{Neu}
\begin{eqnarray}
\label{GF}
\tilde{P}_{\kappa} (x_b \, t_b | x_a \, t_a ) [ j ] = 	
\tilde{P}_{\kappa} (x_b \, t_b | x_a \, t_a ) \exp \left[
\frac{1}{2 D} \int_{t_a}^{t_b} d t_1 x_{\rm cl} ( t_1 ) j ( t_1 ) +
\frac{1}{4 D^2} \int_{t_a}^{t_b} d t_1 \int_{t_a}^{t_b} d t_2
G ( t_1 , t_2 ) j ( t_1 ) j ( t_2 ) \right] \, .
\end{eqnarray}
The quantity $\tilde{P}_{\kappa} (x_b \, t_b | x_a \, t_a )$ is the same as in Eq. (\ref{ABB}), and $x_{\rm cl} ( t_1 )$ 
denotes the classical path
\begin{eqnarray}
x_{\rm cl} ( t_1 ) = \frac{x_a \sinh \kappa ( t_b - t_1 ) + 
x_b \sinh \kappa ( t_1 - t_a )}{\sinh \kappa ( t_b - t_a )} \, .
\end{eqnarray}
The quadratic term in (\ref{GF}) contains the Green function
\begin{eqnarray}
G ( t_1 , t_2 ) = \frac{D
[ \Theta ( t_1 - t_2 ) \sinh \kappa ( t_b - t_1 ) 
\sinh \kappa ( t_2 - t_a ) +  
\Theta ( t_2 - t_1 ) \sinh \kappa ( t_b - t_2 ) 
\sinh \kappa ( t_1 - t_a ) ]}{\kappa \sinh \kappa ( t_b - t_a )}  \, .
\end{eqnarray}
We evaluate harmonic expectation values of polynomials in $x$ 
arising from the generating functional (\ref{GF})
according to a slight generalization of the standard
Wick theorem \cite{KPB,FLORIAN}. The generalization is required by the presence of the linear
term in (\ref{GF}). The evaluation is most economically done in a recursive procedure which we illustrate
with the harmonic expectation value
\begin{eqnarray}
\label{wick}
\langle x^n(\tau_1) \, x^m(\tau_2) \rangle_{\kappa} \, .
\end{eqnarray}
\renewcommand{\labelenumi}{(\roman{enumi})}
\begin{enumerate}
\item Contracting $x (\tau_1)$ with $x^{n-1}(\tau_1)$
and $x^m (\tau_2)$ leads to a
Green function $G (\tau_1, \tau_1)$ and $G (\tau_1, \tau_2)$
with multiplicity
$n-1$ and $m$, respectively. The rest of the factors
remains inside the expectation symbol, leading to
$\langle x^{n-2}(\tau_1) \, x^m (\tau_2) \rangle_{\kappa}$ and
$\langle x^{n-1}(\tau_1) \, x^{m-1} (\tau_2) \rangle_{\kappa}$.
\item If $n>1$, we extract one $x(\tau_1)$ from the expectation value
giving $x_{\rm cl}(\tau_1)$ multiplied
by $ \langle x^{n-1}(\tau_1) x^m (\tau_2) \rangle_{\kappa}$.
\item Add the terms (i) and (ii).
\item Repeat the previous steps 
until only products of expectation values
$ \langle x (\tau_1)\rangle_{\kappa}
= x_{\rm cl} (\tau_1)$ remain. 
\end{enumerate}
With the help of this recursive procedure,
the first-order harmonic expectation value 
$\langle x^4(\tau_1) \rangle_{\kappa}$ is reduced to
\begin{eqnarray}
\label{newwickstheorem}
\langle x^4 (\tau_1) \rangle_{\kappa} 
= x_{\rm cl} (\tau_1) \, \langle x^3 (\tau_1)  \rangle_{\kappa}
+ 3 \, G (\tau_1 , \tau_1) \langle \, x^2 (\tau_1) \rangle_{\kappa} \, .
\end{eqnarray}
Similarly, we find
\begin{eqnarray}
\label{egal}
\langle x^3 (\tau_1) \rangle_{\kappa} 
= x_{\rm cl} (\tau_1) \langle x^2 (\tau_1) 
\rangle_{\kappa} 
+ 2 G (\tau_1, \tau_1) \, x_{\rm cl} (\tau_1) \, ,
\end{eqnarray}
and
\begin{eqnarray}
\label{egal2}
\langle x^2 (\tau_1) \rangle_{\kappa} = x^2_{\rm cl} (\tau_1) 
+ G (\tau_1, \tau_1) \, .
\end{eqnarray}
Combining Eqs. (\ref{newwickstheorem})--(\ref{egal2}), we obtain in first order
\begin{eqnarray}
\label{x4cl}
\langle x^4 (\tau_1) \rangle_{\kappa} 
& = & x^4 _{\rm cl} (\tau_1) \,
+ 6 \, x^2 _{\rm cl} (\tau_1) \, G (\tau_1 , \tau_1) 
+ 3 \, G^2 (\tau_1, \tau_1) \, .
\end{eqnarray}
The contractions can be represented graphically
by Feynman diagrams\index{Feynman diagrams} with the
following rules.
Vertices represent the integrations over $t$
\begin{eqnarray}
\label{vertex}
\parbox{5mm}{\centerline{
\begin{fmfgraph*}(5,5)
\setval
\fmfforce{0w,1/2h}{v1}
\fmfforce{1/2w,1/2h}{v2}
\fmfforce{1w,1/2h}{v3}
\fmf{plain}{v1,v3}
\fmfdot{v2}
\end{fmfgraph*}}}
\hspace*{3mm} \equiv \hspace*{2mm} \int_{t_a}^{t_b} d t \, , \hspace*{2cm}
\parbox{5mm}{\centerline{
\begin{fmfgraph*}(5,5)
\setval
\fmfforce{0w,1h}{v1}
\fmfforce{1w,1h}{v2}
\fmfforce{0w,0h}{v3}
\fmfforce{1w,0h}{v4}
\fmfforce{1/2w,1/2h}{v5}
\fmf{plain}{v1,v4}
\fmf{plain}{v2,v3}
\fmfdot{v5}
\end{fmfgraph*}}}
\hspace*{3mm} \equiv \hspace*{2mm} \int_{t_a}^{t_b} d t \, ,
\end{eqnarray}
a line denotes the Green function\index{Green function}
\begin{eqnarray}
\label{line}
\parbox{10mm}{\centerline{
\begin{fmfgraph*}(10,10)
\setval
\fmfforce{0w,1/2h}{v1}
\fmfforce{1w,1/2h}{v2}
\fmf{plain}{v1,v2}
\fmfv{decor.size=0, label=${\scriptstyle 1}$, l.dist=1mm, l.angle=-180}{v1}
\fmfv{decor.size=0, label=${\scriptstyle 2}$, l.dist=1mm, l.angle=0}{v2}
\end{fmfgraph*}}}
\hspace*{5mm} \equiv \hspace*{2mm} G (t_1, t_2) \, ,
\end{eqnarray}
and a line ending with a cross represents the classical path
\begin{eqnarray}
\label{cross}
\parbox{10mm}{\centerline{
\begin{fmfgraph*}(10,10)
\setval
\fmfforce{0w,1/2h}{v1}
\fmfforce{1w,1/2h}{v2}
\fmf{plain}{v1,v2}
\fmfv{decor.shape=cross,decor.filled=shaded,decor.size=3thick}{v1}
\fmfv{decor.size=0, label=${\scriptstyle 1}$, l.dist=1mm, l.angle=0}{v2}
\end{fmfgraph*}}}
\hspace*{5mm} \equiv \hspace*{2mm} x_{\rm cl} (t_1) \, .
\end{eqnarray}
Inserting the harmonic expectation values (\ref{egal2}) and (\ref{x4cl})  
into the perturbation expansion (\ref{PE}) leads to the
first-order diagrams 
\begin{eqnarray}
\label{PEN}
\tilde{P} ( x_b \, t_b | x_a \, t_a ) = 
\tilde{P}_{\kappa} ( x_b \, t_b | x_a \, t_a ) \left\{
1 + g \left[ \frac{3}{2} \left(
\parbox{10mm}{\centerline{
\begin{fmfgraph*}(10,6)
\setval
\fmfforce{0w,1/2h}{v1}
\fmfforce{1/2w,1/2h}{v3}
\fmfforce{1w,1/2h}{v5}
\fmf{plain}{v1,v5}
\fmfdot{v3}
\fmfv{decor.shape=cross,decor.filled=shaded,decor.size=3thick}{v1}
\fmfv{decor.shape=cross,decor.filled=shaded,decor.size=3thick}{v5}
\end{fmfgraph*}}}
\hspace*{0.2cm} + 
%
\parbox{7mm}{\centerline{
\begin{fmfgraph*}(5,5)
\setval
\fmfforce{1/2w,1h}{v1}
\fmfforce{1/2w,0h}{v3}
\fmf{plain,left=1}{v3,v1,v3}
\fmfdot{v3}
\end{fmfgraph*}}}
\right) - \frac{\kappa}{2 D} \left(
\parbox{10mm}{\centerline{
\begin{fmfgraph*}(10,14)
\setval
\fmfforce{1/2w,12/14h}{v1}
\fmfforce{0w,1/2h}{v2}
\fmfforce{1/2w,1/2h}{v3}
\fmfforce{1w,1/2h}{v4}
\fmfforce{1/2w,2/14h}{v5}
\fmf{plain}{v1,v5}
\fmf{plain}{v2,v4}
\fmfdot{v3}
\fmfv{decor.shape=cross,decor.filled=shaded,decor.size=3thick}{v1}
\fmfv{decor.shape=cross,decor.filled=shaded,decor.size=3thick}{v2}
\fmfv{decor.shape=cross,decor.filled=shaded,decor.size=3thick}{v4}
\fmfv{decor.shape=cross,decor.filled=shaded,decor.size=3thick}{v5}
\end{fmfgraph*}}}
\hspace*{2mm} + 6 \hspace*{2mm}
\parbox{10mm}{\centerline{
\begin{fmfgraph*}(10,10)
\setval
\fmfforce{1/2w,1h}{v1}
\fmfforce{0w,1/2h}{v2}
\fmfforce{1/2w,1/2h}{v3}
\fmfforce{1w,1/2h}{v4}
\fmf{plain}{v2,v4}
\fmf{plain,left=1}{v3,v1,v3}
\fmfdot{v3}
\fmfv{decor.shape=cross,decor.filled=shaded,decor.size=3thick}{v2}
\fmfv{decor.shape=cross,decor.filled=shaded,decor.size=3thick}{v4}
\end{fmfgraph*}}}
\hspace*{2mm} + 3 \hspace*{2mm}
\parbox{10mm}{\centerline{
\begin{fmfgraph*}(10,10)
\setval
\fmfforce{0w,1/2h}{v1}
\fmfforce{1/2w,1/2h}{v2}
\fmfforce{1w,1/2h}{v3}
\fmf{plain,left=1}{v1,v2,v1}
\fmf{plain,left=1}{v2,v3,v2}
\fmfdot{v2}
\end{fmfgraph*}}}
\right) \right] + \ldots \right\}  \, . \label{DIA}
\end{eqnarray}
Evaluating the diagrams, the conditional probability density for the drift coefficient (\ref{LASER}) following from
(\ref{SMF}) and (\ref{DIA}) becomes to the first order in the coupling constant $g$ 
\begin{eqnarray}
P (x_b \, t_b | x_a \, t_a ) &=& P_{\kappa} (x_b \, t_b | x_a \, t_a )
\left\{ 1 + g \left[ c_{00}^{(1)}  ( \tau ) 
+ c_{20}^{(1)} ( \tau ) x_a^2 
+ c_{21}^{(1)} ( \tau ) x_a x_b
+ c_{22}^{(1)} ( \tau ) x_b^2 
\right. \right. \nonumber \\
&&\left. \left. + c_{40}^{(1)}  ( \tau ) x_a^4 
+ c_{41}^{(1)} ( \tau ) x_a^3 x_b
+ c_{42}^{(1)} ( \tau ) x_a^2 x_b^2 
+ c_{43}^{(1)} ( \tau ) x_a x_b^3 
+ c_{44}^{(1)} ( \tau ) x_b^4 
\right] + \ldots \right\} \, ,
\label{FORM}
\end{eqnarray}
where the expansion coefficients $c_{ij}^{(1)} ( \tau )$ are the following functions of the dimensionless variable
$\tau = \kappa ( t_b - t_a )$ (see Fig.~\ref{F1}):
\begin{eqnarray}
c_{00}^{(1)} ( \tau ) & = & \frac{3 D \left[ 1 + 4 e^{- 2 \tau} ( 1 - 2 \tau )
- e^{-4 \tau} (5 + 4 \tau )\right]}{4 \kappa^2( 1 - e^{- 2\tau} )^2} \, ,
\nonumber\\
c_{20}^{(1)} ( \tau ) & = & \frac{3 \left[ e^{-2 \tau} ( 4 \tau -5)
+ 4 e^{- 4 \tau} ( 1 + 2 \tau ) 
+ e^{-6\tau}\right]}{2 \kappa ( 1 - e^{- 2\tau} )^3} = c_{22}^{(1)} ( \tau ) \, ,
\nonumber \\
c_{21}^{(1)} ( \tau ) & = & \frac{3 \left[ e^{-\tau} ( 2- \tau ) 
+ 2 e^{-3 \tau} ( 1- 4 \tau) 
- e^{- 5 \tau} ( 3 \tau + 4 )\right]}{\kappa ( 1 - e^{- 2\tau} )^3} \, ,
\nonumber \\
c_{40}^{(1)} ( \tau ) & = & \frac{2 e^{-2 \tau} + 3 e^{-4 \tau} (1 - 4 \tau)
- 6 e^{- 6 \tau} + e^{-8 \tau}}{4 D ( 1 - e^{- 2\tau} )^4}\, ,
\nonumber\\
c_{41}^{(1)} ( \tau ) & = & \frac{- e^{- \tau} + 3 e^{-3 \tau} ( 4 \tau - 3)
+ 3 e^{-5 \tau} ( 3 + 4 \tau )+ e^{- 7 \tau}}{2 D ( 1 - e^{- 2\tau} )^4}
= c_{43}^{(1)} (\tau )\, ,
\nonumber \\
c_{42}^{(1)} ( \tau ) & = & \frac{3 \left[ e^{-2 \tau} ( 3 - 2 \tau) 
-8 \tau e^{-4 \tau} 
- e^{- 6 \tau} ( 3 + 2 \tau)\right]}{2 D ( 1 - e^{- 2\tau} )^4}\, ,
\nonumber \\
c_{44}^{(1)} ( \tau ) & = & \frac{-1 + 6 e^{-2 \tau} - 3 e^{-4 \tau}
( 1 + 4 \tau) - 2 e^{-6 \tau}}{4 D ( 1 - e^{- 2\tau} )^4}\, . \label{COEFF}
\end{eqnarray}
It is easy to verify that the conditional probability density (\ref{FORM})--(\ref{COEFF}) 
\renewcommand{\labelenumi}{(\roman{enumi})}
\begin{enumerate}
\item obeys the corresponding Fokker-Planck
equation following from (\ref{FPE}), (\ref{FPO}) and (\ref{LASER})
\begin{eqnarray}
\frac{\partial}{\partial t_b} P ( x_b \, t_b | x_a \, t_a )
= \frac{\partial}{\partial x_b} \left[ \left( \kappa x_b + g x_b^3 \right)  
P ( x_b \, t_b | x_a \, t_a ) \right] + D \frac{\partial^2}{\partial x_b^2}
P ( x_b \, t_b | x_a \, t_a ) \, ,
\end{eqnarray}
with the initial condition (\ref{IVP}), because of 
\begin{eqnarray}
c_{00}^{(1)} (0) = c_{20}^{(1)} (0) = c_{21}^{(1)} (0) = c_{22}^{(1)} (0) = c_{41}^{(1)} (0) = c_{42}^{(1)} (0) = 
c_{43}^{(1)} ( 0 ) = 0 \, , \quad c_{40}^{(1)} ( 0 ) = - c_{44}^{(1)} ( 0 ) = \frac{1}{8 D} \, .
\end{eqnarray} 
\item is normalized according to (\ref{NORM}) for all times $t_b$.
\item approaches the stationary solution (\ref{NLST}) in the long-time limit (\ref{SS1}).
\end{enumerate}
All these properties of $P ( x_b \, t_b | x_a \, t_a )$ are satisfied to first order in the coupling constant $g$. As such
the perturbative result (\ref{FORM})--(\ref{COEFF}) 
is only applicable for small values of the coupling constant $g$. This limitation is removed by a variational evaluation
which enables us to find an approximate conditional probability density for all values of the coupling constant $g$.
\section{Variational Perturbation Theory}
Variational perturbation theory in quantum statistics approximates a given potential by adding and subtracting a local trial
harmonic oscillator whose frequency is then systematically optimized \cite{Kleinert}. Here we transfer this 
procedure to Markov processes and modify the nonlinear drift coefficient (\ref{LASER}) by
adding and subtracting a trial Brownian motion with a yet unknown damping constant $K$:
\begin{eqnarray}
K ( x ) = - K x - g \left( \frac{\kappa-K}{g} x + x^3 \right) \, .
\end{eqnarray}
After this, we treat the combined second term as being small of the order of the coupling constant $g$. The result is obtained
most simply by replacing the damping constant $\kappa$ in the original drift coefficient (\ref{LASER}) according to
the substitution rule (compare with Chap.~5 in Ref.~\cite{Kleinert})
\begin{eqnarray}
\label{TRICK}
\kappa \quad \rightarrow \quad K \left( 1 + gr \right) \, ,
\end{eqnarray}
where we have introduced the abbreviation
\begin{eqnarray}
\label{ABBR}
r=\frac{\kappa - K}{g K} \, .
\end{eqnarray}
Writing the conditional probability density (\ref{FORM}) as
\begin{eqnarray}
\label{REWR}
P ( x_b \, t_b | x_a \, t_a ) = \exp \left[ W  ( x_b \, t_b ; x_a \, t_a ) \right] \, ,
\end{eqnarray}
we apply the  substitution rule (\ref{TRICK}) to the cumulant expansion $W  ( x_b \, t_b ; x_a \, t_a )$, and reexpand up to the
first order in the coupling constant $g$. Afterwards the abbreviation $r$ is reexpressed in terms of $\kappa$ and $K$
via (\ref{ABBR}). This leads to
\begin{eqnarray}
W^{(1)} (x_b \, t_b ; x_a \, t_a ; K ) &=& \left\{ c_{00}^{(0)} ( \tau ) + c_{20}^{(0)} ( \tau ) x_a^2 
+ c_{21}^{(0)} ( \tau ) x_a x_b
+ c_{22}^{(0)} ( \tau ) x_b^2 
+ g \left[ c_{00}^{(1)} ( \tau ) 
+ c_{20}^{(1)} ( \tau ) x_a^2 
+ c_{21}^{(1)} ( \tau ) x_a x_b
\right. \right. \nonumber \\
&&\left. \left. + c_{22}^{(1)} ( \tau ) x_b^2 
+ c_{40}^{(1)} ( \tau ) x_a^4 
+ c_{41}^{(1)} ( \tau ) x_a^3 x_b
+ c_{42}^{(1)} ( \tau ) x_a^2 x_b^2 
+ c_{43}^{(1)} ( \tau ) x_a x_b^3 
+ c_{44}^{(1)} ( \tau ) x_b^4 
\right] \right\} \, ,
\label{VAR}
\end{eqnarray}
where the zeroth-order expansion coefficients 
\begin{eqnarray}
c_{00}^{(0)} ( \tau ) &=& \frac{1}{2} \ln \frac{K}{2 \pi D (1-e^{-2\tau})} + \left(\frac{\kappa}{K} -1 \right)
\left( \frac{1}{2} -  \frac{\tau e^{-2\tau}}{1-e^{-2\tau}} \right) \, , \nonumber\\
c_{20}^{(0)} ( \tau ) &=& - \frac{\kappa e^{-2\tau}}{2 D (1-e^{-2\tau})}
+ \frac{(\kappa - K) \tau e^{-2\tau}}{D (1-e^{-2\tau})^2} \, , \nonumber\\
c_{21}^{(0)} ( \tau ) &=& \frac{\kappa e^{-\tau}}{D (1-e^{-2\tau})} 
- \frac{(\kappa - K) \tau (1 + e^{-2\tau})e^{-\tau}}{D (1-e^{-2\tau})^2} \, , \nonumber\\
c_{22}^{(0)} ( \tau ) &=& - \frac{\kappa}{2 D (1-e^{-2\tau})}
+ \frac{(\kappa - K)\tau e^{-2\tau}}{D (1-e^{-2\tau})^2}
\end{eqnarray}
and the first-order expansion coefficients (\ref{COEFF}) are functions of the dimensionless variable
$\tau = K ( t_b - t_a )$. Note that using (\ref{COEFF}) in (\ref{VAR}) necessitates to substitute $\kappa$ by $K$.\\

We now remove the dependence of 
the cumulant expansion (\ref{VAR}) from the artificially introduced trial damping constant $K$. According to the
principle of minimal sensitivity \cite{STEVENSON}, we minimize its influence on $W^{(1)} (x_b \, t_b | x_a \, t_a ; K )$
by searching for local extrema of $W^{(1)} (x_b \, t_b | x_a \, t_a ; K )$ with respect to $K$, i.e. from the condition
\begin{eqnarray}
\label{COND}
\left. \frac{\partial W^{(1)} (x_b \, t_b ; x_a \, t_a ; K )}{\partial K}\right|_{K=K^{(1)} (x_b \, t_b ; x_a \, t_a )} = 0 \, .
\end{eqnarray}
It may happen that this equation is not solvable within a certain region of the parameters $x_b, t_b, x_a, t_a$. In this
case we look for turning points instead in accordance with the principle of minimal sensitivity \cite{Kleinert,FLORIAN}, i.e.
we determine the variational parameter $K^{(1)} (x_b \, t_b ; x_a \, t_a )$ from solving
\begin{eqnarray}
\label{CONDB}
\left. \frac{\partial^2 W^{(1)} (x_b \, t_b ; x_a \, t_a ; K )}{\partial K^2}\right|_{K=K^{(1)} (x_b \, t_b ; x_a \, t_a )} = 0 \, .
\end{eqnarray}
The solution $K^{(1)} (x_b \, t_b ; x_a \, t_a )$ from (\ref{COND}) or (\ref{CONDB}) yields the variational result
\begin{eqnarray}
\label{SOLV}
P ( x_b \, t_b | x_a \, t_a ) \approx 
\frac{\exp \left[ W  \left( x_b \, t_b ; x_a \, t_a ; K^{(1)} (x_b \, t_b ; x_a \, t_a ) 
\right)\right]}{\displaystyle \int_{-\infty}^{+ \infty} d x_b \exp \left[ W  \left( x_b \, t_b ; x_a \, t_a ; 
K^{(1)} (x_b \, t_b ; x_a \, t_a ) \right)\right]}
\end{eqnarray}
for the conditional probability density. Note that variational perturbation theory does not preserve the normalization of the
conditional probability density. Although the perturbative result (\ref{FORM})--(\ref{COEFF}) is still normalized in the
usual sense (\ref{NORM}) to first order in the coupling constant $g$, 
this normalization is spoilt by choosing an $x_b$-dependent damping constant $K^{(1)}  ( x_b \, t_b ; x_a \, t_a)$.
Thus we have to normalize the variational conditional probability density according to (\ref{SOLV}) 
at the end (compare the similar situation for the variational ground-state wave function in Refs. \cite{FLORIAN,KUNIHIRO}).\\

\begin{figure}[h]
\centerline{\includegraphics[width=14cm]{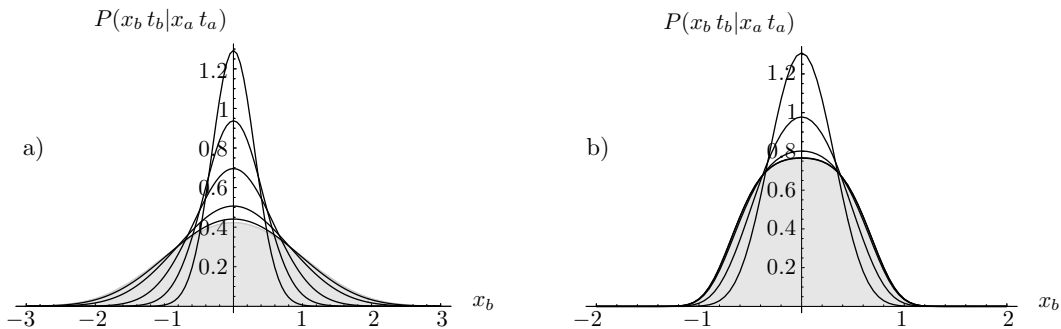}}
\caption{\label{F2}Conditional probability density $P (x_b\, t_b | x_a\, t_a)$
below the laser threshold at $\kappa=1$, $D=1$, and $x_a=t_a=0$. For
the coupling constants $g=0.1$ in a) and $g=10$ in b) the distribution is shown for the times $t_b = 0.05,0.1,0.2,0.5,1$
from the top to the bottom at the origin, respectively.}
\end{figure}

\begin{figure}[h]
\centerline{\includegraphics[width=14cm]{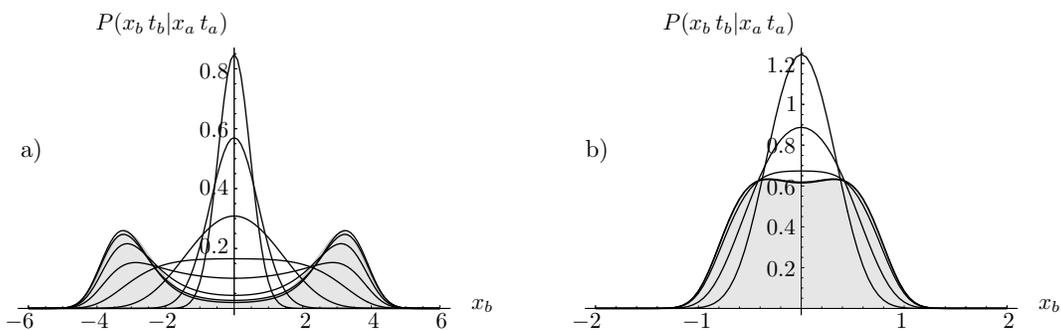}}
\caption{\label{F3}Conditional probability density $P (x_b\, t_b | x_a\, t_a)$ 
above the laser threshold at $\kappa=-1$, $D=1$, and $x_a=t_a=0$. For
the coupling constants $g=0.1$ in a) and $g=10$ in b) the distribution is shown for the times $t_b = 0.1,0.2,0.5,1,1.5,2,3,4$
and $t_b=0.05,0.1,0.2,0.5$ from the top to the bottom at the origin, respectively.}
\end{figure}

We have applied variational perturbation theory to analyze the nonlinear stochastic model (\ref{LASER}) 
with $D = 1, x_a = t_a = 0$ below and above the laser threshold, i.e. $\kappa = 1$
and $\kappa =-1$, for small and strong coupling $g=0.1$ and $g=10$. 
The results for the conditional probability density $P (x_b\, t_a | x_a\, t_a)$ 
are plotted in Figs.~\ref{F2} and \ref{F3}. 
On the scale of the figures, they show no significant
deviation from numerical solutions of the corresponding Fokker-Planck equation, some minor deviations
only occur above the laser threshold $\kappa = -1$ for $g=0.1$ \cite{DREGER}. 
Both figures illustrate how the densities, originally peaked at the origin,
turn into their stationary solutions in the long-time limit. The stationary solution reveals below the threshold 
one extremum in Fig.~\ref{F2}, whereas above the laser threshold in Fig.~\ref{F3} two extrema occur.
\section{Conclusion}
We have presented the lowest-order variational calculation for the conditional probability density of the nonlinear drift 
coefficient (\ref{LASER}). By going to higher orders in variational perturbation theory, it is straightforward to increase
systematically the accuracy to any desired degree \cite{Kleinert}.
\section{Acknowledgement}
The authors thank Jens Dreger for preparing the final Figs. 2 and 3, Dr. Anna Okopi\'nska for
carefully reading the manuscript, and Dr. Bodo Hamprecht for many discussions.
M.V.P. is grateful for receiving a DAAD fellowship during which
the research for this paper was performed.
\section*{Note Added in Proof}
Meantime we became aware of the alternative approach \cite{ANNA2} to variationally analyze the nonlinear model (\ref{LASER}).
Above the laser threshold $\kappa = -1$, that method
yields for all times $t_b$ a unique solution of the extremal condition corresponding to (\ref{COND}). However,
the resulting conditional probability density shows for larger times $t_b$ significant deviations from our and from numerical 
solutions of the Fokker-Planck equation. 
\end{fmffile}
\end{document}